\documentclass[aps,preprintnumbers,eqsecnum,amsmath,amssymb,showpacs,nofootinbib]{revtex4}
\usepackage{graphics}
\usepackage{dcolumn}
\usepackage{bm}
\usepackage{epsfig}

\begin{document} 

\title{\boldmath Local-duality QCD sum rules for strong isospin breaking \\
in the decay constants of heavy-light mesons}
\author{Wolfgang Lucha$^a$, Dmitri Melikhov$^{a,b,c}$, and Silvano Simula$^d$}
\affiliation{$^a$Institute for High Energy Physics, Austrian Academy of Sciences, Nikolsdorfergasse 18, A-1050 Vienna, Austria\\ 
$^b$D.~V.~Skobeltsyn Institute of Nuclear Physics, M.~V.~Lomonosov Moscow State University, 119991, Moscow, Russia\\
$^c$Faculty of Physics, University of Vienna, Boltzmanngasse 5, A-1090 Vienna, Austria\\ 
$^d$Istituto Nazionale di Fisica Nucleare, Sezione di Roma Tre, Via della Vasca Navale 84, I-00146 Roma, Italy}

\begin{abstract}
We discuss the leptonic decay constants of heavy--light mesons by
means of Borel QCD sum~rules in the local-duality (LD) limit of
infinitely large Borel mass parameter. In this limit, for an
appropriate choice of the invariant structures in the QCD
correlation functions, all vacuum-condensate contributions vanish
and all nonperturbative effects are contained in only one
quantity, the effective threshold. We study properties of the LD
effective thresholds in the limits of large heavy-quark mass $m_Q$ and
small light-quark mass $m_q$. In the heavy-quark limit, we clarify the role 
played by the radiative corrections in the effective threshold for reproducing 
the pQCD expansion of the decay constants of pseudoscalar and vector mesons. 
We show that the dependence of the meson decay constants on $m_q$ arises predominantly 
(at the level of 70--80\%) from the calculable $m_q$-dependence of the
perturbative spectral densities. Making use of the lattice QCD
results for the decay constants of nonstrange and strange pseudoscalar and vector heavy
mesons, we obtain solid predictions for the decay constants of
heavy--light mesons as functions of $m_q$ in the range from a few
to 100 MeV and evaluate the corresponding strong isospin-breaking effects:  
$f_{D^+} - f_{D^0}=(0.96 \pm 0.09) ~ {\rm MeV}$, 
$f_{D^{*+}} - f_{D^{*0}}= (1.18 \pm 0.35) ~ {\rm MeV}$, 
$f_{B^0} - f_{B^+}=(1.01 \pm 0.10) ~ {\rm MeV}$, 
$f_{B^{*0}} - f_{B^{*+}}=(0.89 \pm 0.30) ~ {\rm MeV}$.
\end{abstract}


\maketitle

\section{Introduction}

The method of QCD sum rules \cite{svz}, based on
the exploitation of Wilson's operator product expansion (OPE) in
the study of properties of individual hadrons, has been
extensively applied to the decay constants of heavy mesons \cite{aliev,rubinstein}.
An important finding of these analyses was the observation of the
strong sensitivity of the decay constants to the precise values of
the input OPE parameters and to the algorithm used for fixing the
effective threshold \cite{lms_1}. 
For any given approximation of the hadronic spectral density based 
on quark-hadron duality, the effective threshold determines to a large 
extent the numerical prediction for the decay constants inferred from QCD sum rules:  
even if the parameters of the truncated OPE are known with high precision, 
the decay constants may be predicted with only a limited accuracy, which 
represents their systematic uncertainty. In a series of
papers \cite{lms_new}, we proposed a new algorithm for fixing the
effective threshold within the Borel QCD sum rules which
allowed us to obtain realistic estimates of the systematic
uncertainties. Our procedure opened the possibility to get
predictions for the decay constants with a controlled accuracy
\cite{lms_fp1,lms_fp2} 
and thus allowed us to
address subtle effects that call for a profound accurate
treatment, such as the ratios of the decay constants of heavy
vector and pseudoscalar mesons \cite{lms_fB_ratio} or the strong
isospin-breaking (IB) effects in the decay constants of heavy-light
mesons~\cite{lms_ib_2017}, generated by the mass difference ($m_d - m_u$) between up and down quarks. 

Here, we discuss the application of another variant of QCD sum
rules to the evaluation of the strong IB effects in the decay constants of heavy-light
pseudoscalar and vector mesons.
Our analysis takes advantage of
the fact that the OPE provides the analytic dependence of the
correlation functions on the quark masses; this allows us to
study, e.g., the impact of the light-quark mass on heavy-meson
decay constants, thus providing access to the strong IB effects. The approach we describe in this work seems
quite promising for studying the dependence of a generic hadron
observable on quark masses.

\vspace{-0.05cm}
\subsection{QCD sum rule in the local-duality limit}

A typical Borel QCD sum rule for the decay constant $f_H$ of a
heavy (pseudoscalar or vector) $\bar Qq$ meson $H$ of mass $M_H$,
consisting of a heavy quark $Q$ with mass $m_Q$ and a light quark
$q$ with mass $m_q$, has the form
\begin{eqnarray}
\label{srN}
f_H^2 (M_H^2)^N e^{-M_H^2 \tau}=\int\limits_{(m_Q+m_q)^2}^{s^{(N)}_{\rm eff}(\tau,m_Q,m_q,\alpha_s)}ds e^{-s \tau}s^N\rho_{\rm pert}(s,m_Q,m_q,\alpha_s) 
+ \Pi^{(N)}_{\rm power}(\tau,m_Q,m_q,\alpha_s,\langle \bar qq\rangle,...). ~
\end{eqnarray}
Here, $\tau $ is the Borel parameter, $N$ is an integer that
depends on the Lorentz structure in the correlation function
chosen for the sum rule and on the number of subtractions in the
corresponding dispersion representation, and $s_{\rm eff}$ is the effective 
threshold such that $\sqrt{s_{\rm eff}}$ lies between the mass of the ground-state and the first excited state \cite{svz}, namely
$s_{\rm eff}=(M_H + z_{\rm eff})^2$ with $z_{\rm eff} \simeq 0.4$-$0.5$ GeV. 

Nonperturbative effects appear on the r.h.s.~of (\ref{srN}) at two places: as power
corrections given in terms of vacuum condensates and in the
effective threshold $s^{(N)}_{\rm eff}$. Depending
on the chosen value of $N$, nonperturbative effects are
distributed~in a different way between power corrections and the
effective threshold. Perturbative effects are encoded in the spectral density $\rho_{\rm pert}$,
in the effective threshold $s^{(N)}_{\rm eff}$ and in the power corrections 
$\Pi^{(N)}_{\rm power}$.

Recall that Eq.~(\ref{srN}) is based on modelling 
the hadron continuum as the effective continuum, i.e., on the substitution $\rho_{\rm cont}(s) = \theta(s - s_{\rm eff})\rho_{\rm pert}(s)$. 
This relation is fulfilled point-wise at large values of $s$ above some $s_{\rm up}$, but is a ``weak'' relation and requires an 
appropriate smearing for $s$ in the mid-energy region above the physical hadron continuum 
threshold $s_{\rm th}$. An appropriate smearing is reached by performing the Borel transform 
\begin{eqnarray}
\label{duality}
\int_{s_{\rm th}}^{s_{\rm up}} ds \rho_{\rm hadr}(s) e^{-s \tau} = \int_{s_{\rm eff}}^{s_{\rm up}} ds \rho_{\rm pQCD}(s) e^{-s \tau} . 
\end{eqnarray}
For nonzero $\tau$, the contribution of the hadron continuum given by (\ref{duality}) 
is exponentially suppressed compared to the ground-state contribution (\ref{srN}).
Therefore, in the conventional use of QCD sum rules one works in some window of nonzero values of $\tau$. 
However, one may ask whether or not the relation (\ref{srN}) may be extended down to $\tau=0$, the so-called local-duality (LD) 
limit\footnote{The LD limit in 
Borel sum rules was introduced and discussed in \cite{LD,ld1} in connection with the pion and 
nucleon elastic form factors, and later applied to the analysis of meson transition form factors in \cite{lmld}. 
A specific feature of this limit is the vanishing of power corrections in the two- and three-point Borelized correlation 
functions of axial-vector and vector currents of light quarks.}.
Obviously, at $\tau=0$ an appropriate smearing in (\ref{duality}) is guaranteed by the integration; 
on the other hand, the excited states are not suppressed and one can doubt that modelling the hadron continuum as the 
effective continuum remains a good approximation at small $\tau$.

First, note that power corrections contain singular terms of the form $\tau^{2-N}\log(\tau)$. 
Therefore, the limit $\tau \to 0$ cannot be easily taken in the sum rule (\ref{srN}) for $N\ge 2$. For $N=0$ and $N=1$, 
the limit $\tau\to 0$ in (\ref{srN}) is mathematically well defined. To demonstrate that this limit is also physically meaningful, 
one needs to show that the corresponding $s_{\rm eff}$ indeed lies in the expected range. 

\begin{figure}[htb!]
\begin{tabular}{cc}
\includegraphics[width=8.0cm]{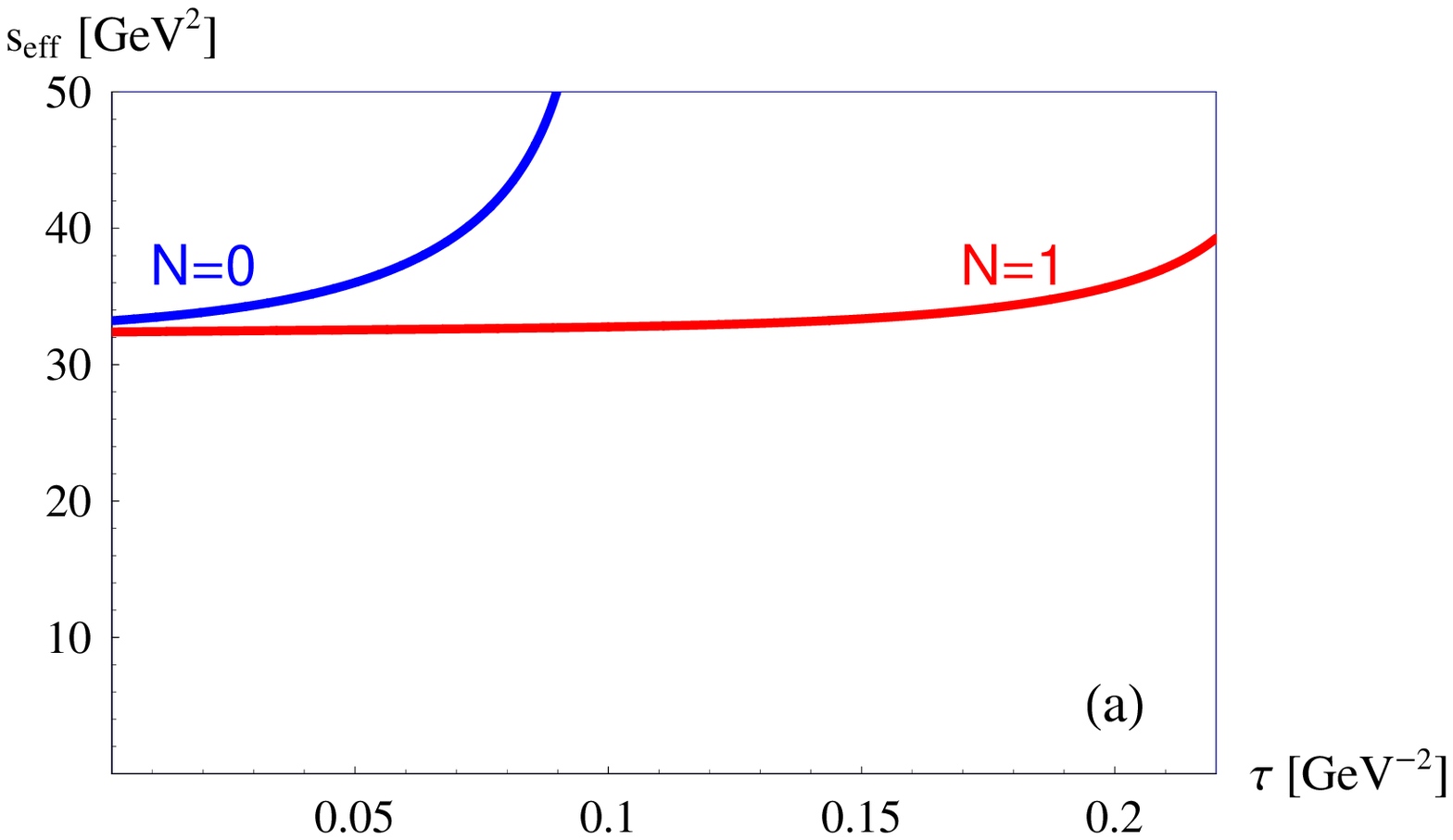} & 
\includegraphics[width=8.0cm]{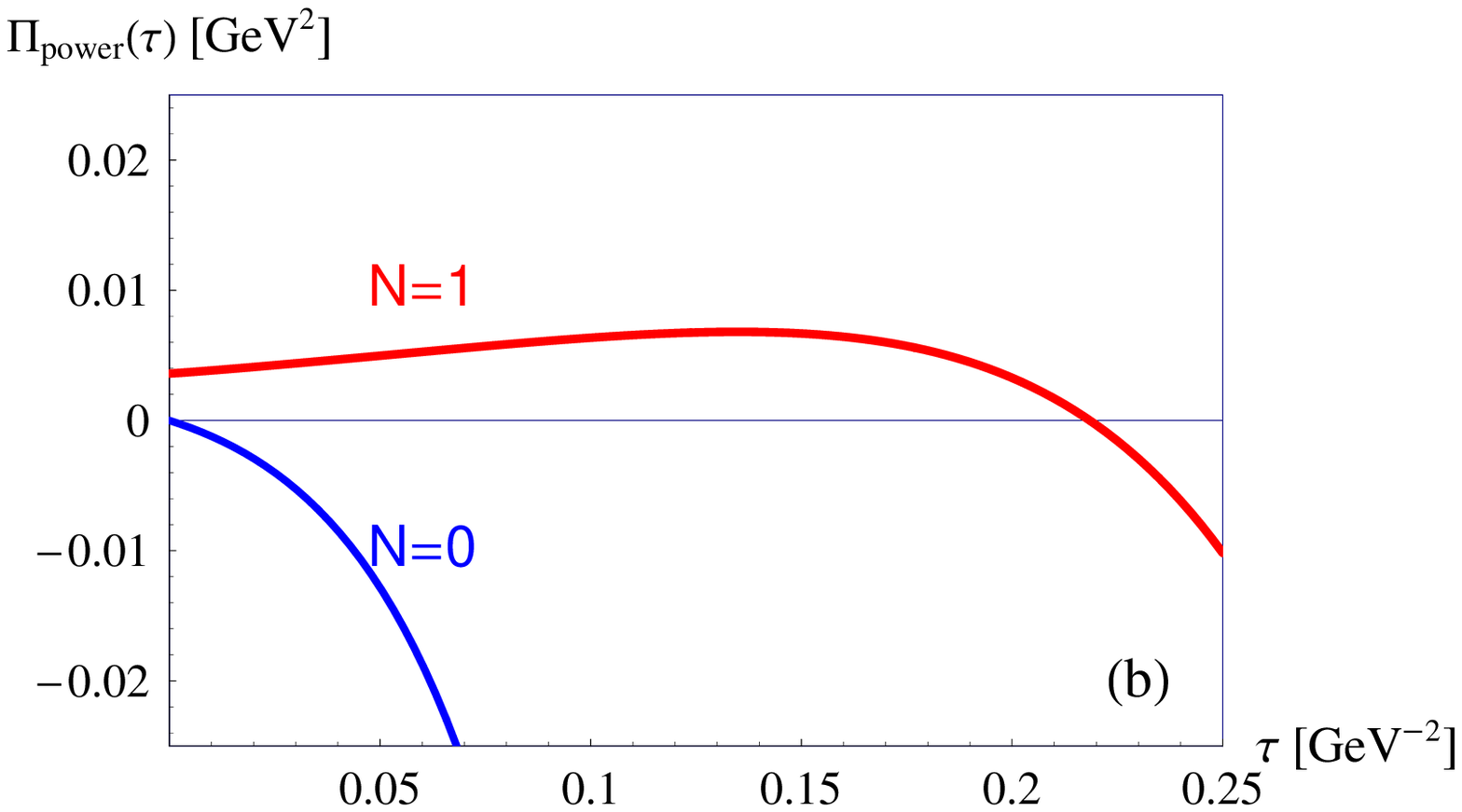}
\end{tabular}\caption{\label{Plot:seff} (a) The effective threshold obtained by solving Eq.~(\ref{srN}) in the case of the $B^*$-meson for $N=0$ and $N=1$, using in the r.h.s.~the leading order (LO), NLO and NNLO perturbative contributions and the condensates up to dimension 6, while the l.h.s.~is calculated adopting the known values for $f_{B^*}$~\cite{Lubicz:2017asp} and $M_{B^*}$~\cite{pdg}. (b) Truncated power corrections including condensates up to dimension 6 for $N=0$ (see Sect.~\ref{OPE}) and for $N=1$~\cite{khod}.}
\end{figure}

Figure~\ref{Plot:seff}(a) presents the effective threshold obtained in the case of the vector $B^*$-meson by solving Eq.~(\ref{srN}) for $N=0$ and $N=1$, 
using in its l.h.s.~the results of recent lattice QCD simulations for $f_{B^*}$~\cite{Lubicz:2017asp} and the experimental value for $M_{B^*}$~\cite{pdg}. 
Figure~\ref{Plot:seff}(b) shows the truncated series of power corrections 
including operators of dimension up to 6 for both $N=0$ and $N=1$. Note that 
power corrections at $\tau=0$ vanish for $N=0$ and take a finite value for $N=1$. The vanishing of $N=0$ power corrections 
at $\tau=0$ is related to the absence in QCD of a dimension-2 condensate. 
Obviously, the truncated power corrections for $N=1$ remain under control in a rather broad range of $\tau$, but 
for $N=0$ explode relatively soon as $\tau$ increases. 

It is clear that for the $N=1$ case the OPE is under good control in a broad range of $\tau$ and therefore the 
lower boundary of the Borel window can be safely extended down to $\tau=0$. 
For $N=0$ only at relatively small $\tau$ the OPE is under control and the approximation of a $\tau$-independent effective threshold may work well.
The relevance of the unknown higher-order power corrections is reflected by the sharp rise of the effective threshold visible in Fig.~\ref{Plot:seff}(a). 
Obviously, the standard QCD sum-rule analysis based on a stability window with a constant effective threshold may be problematic. 
In this respect an alternative approach based on a $\tau$-dependent effective threshold seems to be more appropriate, but this issue goes 
well beyond the scope of the present paper.
Here, the only important property is that the effective threshold at $\tau=0$ has the value expected on the basis of the standard 
considerations \cite{svz}, i.e., it is around $(M_{B^*} + z_{\rm eff})^2$ with $z_{\rm eff} \sim 0.4$-$0.5$ GeV. 
Therefore, for the correlators with $N<2$, modelling the hadron continuum as an effective continuum remains a valid 
and equally accurate approximation as $\tau \to 0$, which does not represent a point of discontinuity.  

In this work we show that the sum rule (\ref{srN}) for $N=0$ can be of particular interest. 
Obviously, considering the sum rule at only one point, $\tau=0$, does not allow for the use of the 
usual sum-rule stability criteria \cite{svz} for determining $s_{\rm eff}$. Consequently, the decay constants 
cannot be determined entirely within the QCD sum rules; some ``external'' inputs are needed to determine $s_{\rm eff}$. 
Nevertheless, in this work we will show that, besides the reduction of the uncertainties related to the absence of the 
condensates, the LD sum rules represent an efficient tool to investigate the dependence of the pseudoscalar and vector 
meson decay constants on quark masses and their perturbative behavior in QCD.
Moreover, the LD sum rules turn out to be particularly suitable for the analysis of the strong IB effects in the two-point functions 
when implemented with only few ``external'' inputs, e.g., from experiment or lattice QCD.

\subsection{Strong isospin breaking from a QCD sum rule in the LD limit}

We are interested in the dependence of the decay constants of
heavy--light mesons on the quark masses, in particular, in the
strong IB effects in the decay constants (i.e., the
difference between the decay constants of $\bar Qd$~and $\bar Qu$
mesons induced by the small mass difference $\delta m=m_d-m_u$).
We therefore need to properly take into account all effects
depending on the light-quark flavour $q$ in the correlation
function of the appropriate $\bar Q q$ interpolating~currents.

Clearly, the $m_q$-dependence on the l.h.s.\ of (\ref{srN}) is
encoded both in the decay constant $f_H$ and in the meson
mass~$M_H$. On the r.h.s, the IB effects come
from several sources: the $m_q$-dependence of $\rho_{\rm pert}(s,m_Q,m_q,\alpha_s)$, the $m_q$-dependence of the effective
threshold $s_{\rm eff}$, the $m_q$-dependence of the power corrections, and the flavour dependence of the quark condensates,
in particular, of $\langle \bar qq\rangle$. In general, all these
effects mix together, which renders~the goal of isolating the IB effects in $f_H$ a complicated task. 
A careful analysis has been carried out recently in \cite{lms_ib_2017}, following the standard choice $N=2$ 
for pseudoscalar and $N=1$ for vector mesons. 

There is, however, a special case which makes the sum rule (\ref{srN}) particularly suitable 
for the analysis of IB effects. As discussed above, for $N=0$ and $N=1$ power corrections are regular functions at $\tau=0$. 
Moreover, for $N=0$ power corrections at $\tau=0$ vanish (power corrections for $N=1$ are nonzero at $\tau=0$) 
and Eq.~(\ref{srN}) is reduced to 
\begin{eqnarray}
\label{sr0}
f_H^2=\int\limits_{(m_Q+m_q)^2}^{s_{\rm eff}(m_Q,m_q,\alpha_s)}ds \rho_{\rm pert}(s,m_Q,m_q,\alpha_s).
\end{eqnarray}
On the l.h.s., the $M_H$ contribution has dropped out, thus
opening a direct access to the $m_q$ dependence of the decay
constant $f_H$. Since power corrections do not contribute to the sum rule in the LD limit, all
nonperturbative effects enter now through a single quantity -- the effective threshold. 
The functional dependence of the perturbative spectral density on the quark masses $m_Q$ and $m_q$ 
can be calculated to the necessary accuracy. The functional dependence of $s_{\rm eff}$ on the quark masses may be
determined from the general properties of the decay constants of heavy--light mesons in QCD. 
Namely, its dependence on $m_q$ may be parameterized by a polynomial formula in $m_q$ plus 
chiral logs, which can be determined by matching to heavy-meson chiral perturbation theory. 
The numerical coefficients in the polynomial function are not known but may be determined using only few results 
on the decay constants from lattice QCD, e.g., for
nonstrange and strange heavy mesons. Having at our disposal the explicit $m_q$ dependence of the effective threshold
and of the spectral densities opens direct access to 
the strong IB effects related to the small
difference of the $u$- and $d$-quark masses in QCD.

We will demonstrate that the main $m_q$ dependence of the decay constants originates from
the calculable $m_q$ dependence of the perturbative spectral densities. Therefore, the LD limit opens the possibility of a
reliable analysis of the $m_q$ dependence and the strong IB effects in the decay constants of heavy--light mesons 
(and, in principle, also in other quantities).

This paper is organized as follows: In Sec.~2, we recall the
spectral densities of the QCD correlation functions relevant for
our LD sum-rule analysis. In Sec.~3, we study the $m_Q$- and
$m_q$-dependences of the effective thresholds by making use of an
appropriate mass scheme (pole mass and running mass) for the heavy
quarks. In Sec.~4, we perform the numerical analysis of the decay
constants of heavy--light pseudoscalar and vector mesons and
obtain predictions for strong IB effects in the decay constants.
Section 5 gives our conclusions. The Appendix \ref{sec:appendix} collects some
details of treating the IB effects within the OPE, 
which, in our opinion, deserve to be presented.

\section{\label{OPE} Local-duality sum rules for $f_P$ and $f_V$}

Let us consider two-point QCD sum-rules for decay constants of
pseudoscalar ($P$) and vector ($V$) mesons built up of one massive
quark $Q$ with mass $m_Q$ and one light quark $q$ with mass $m_q$.
We consider the axial-vector current
\begin{eqnarray}
\label{A}
j_\mu^5(x)=\bar q(x) \gamma_\mu \gamma_5 Q(x)
\end{eqnarray}
and the vector current
\begin{eqnarray}
\label{V}
j_\mu(x)=\bar q(x) \gamma_\mu Q(x)
\end{eqnarray}
as interpolating currents for the pseudoscalar and vector mesons,
respectively. The corresponding correlation functions involve two
Lorentz structures, the transverse structure $g_{\mu\nu}p^2-p_\mu
p_\nu$ and the longitudinal structure $p_\mu p_\nu$:
\begin{eqnarray}
\label{pi5munu} \Pi^{5}_{\mu\nu}(p)&=&i\int dx e^{i p x} \langle
T( j_\mu^{5}(x){j^{5}_\nu}^{\dagger}(0))\rangle = \left(g_{\mu\nu}p^2-p_\mu p_\nu\right)\Pi^{5}_{\rm T}(p^2)+p_\mu
p_\nu\Pi^{5}_{\rm L}(p^2),
\\
\label{pimunu} \Pi_{\mu\nu}(p)&=&i\int dx e^{i p x} \langle T(
j_\mu(x){j_\nu}^{\dagger}(0))\rangle = \left(g_{\mu\nu}p^2-p_\mu p_\nu\right)\Pi_{\rm T}(p^2)+p_\mu p_\nu\Pi_{\rm L}(p^2) ~ . \quad
\end{eqnarray}
For $\Pi^5_{\mu\nu}(p)$, we study the longitudinal structure
$p_\mu p_\nu$, as it contains the ground-state pseudoscalar-meson
contribution
\begin{eqnarray}
p_\mu p_\nu \frac{f^2_{P}}{M_{P}^2-p^2},
\end{eqnarray}
with
\begin{eqnarray}
\langle 0|\bar q \gamma_\mu \gamma_5 Q|P(p)\rangle=i f_{P}p_\mu.
\end{eqnarray}
For $\Pi_{\mu\nu}(p)$, we study the transverse structure $g_{\mu \nu}p^2-p_\mu p_\nu$, which contains the ground-state vector-meson
contribution
\begin{eqnarray}
(g_{\mu\nu}M_V^2-p_\mu p_\nu)\frac{f^2_{V}}{M_{V}^2-p^2},
\end{eqnarray}
with
\begin{eqnarray}
\langle 0|\bar q \gamma_\mu Q|V(p)\rangle=M_{V}f_{V}\epsilon_\mu(p).
\end{eqnarray}
As already noted, power corrections for dimension-2 correlation
functions vanish in the LD limit $\tau=0$. 

We now present
their explicit form. The leading power correction to
$\Pi^5_{\mu\nu}(p)$ is given by the quark condensate and
easily~derived:
\begin{eqnarray}
\Pi^5_{\mu\nu}(p)|_{\langle \bar qq\rangle}
&=&
-\left(p^2 g_{\mu\nu}-p_\mu p_\nu\right)m_Q\langle \bar qq\rangle 
\left[\frac{1}{p^2}\left(\frac{1}{m_Q^2-p^2}+\frac{\frac12 m_Q m}{(m_Q^2-p^2)^2}\right)\right]\nonumber\\
&+&p_\mu p_\nu\langle \bar qq\rangle
\left[\frac{-m_Q}{p^2}\left(\frac{1}{m_Q^2-p^2}+\frac{\frac12 m_Q m}{(m_Q^2-p^2)^2}\right) + \frac{m}{(m_Q^2-p^2)^2}\right] ~ .
\end{eqnarray}
The Borel transform $p^2\to\tau$ [defined such that
$\frac{1}{a-p^2}\to e^{-a \tau}$] of $\Pi^5_{\mu\nu}(p,q=0)|_{\langle \bar qq\rangle}$ reads
\begin{eqnarray}
&-&\left(p^2 g_{\mu\nu}-p_\mu p_\nu\right)\frac{\langle \bar qq\rangle}{m_Q} \left[1-e^{-m_Q^2\tau} +\frac{m}{2m_Q}\left(1-e^{-m_Q^2\tau}-\tau m_Q^2 e^{-m_Q^2\tau}\right)\right]\nonumber\\
&+&p_\mu p_\nu\frac{\langle \bar qq\rangle}{2m_Q^2}\left[-(2m_Q+m)\left(e^{-m_Q^2\tau}-1\right) +m\, m_Q^2\tau e^{-m_Q^2\tau}\right].
\end{eqnarray}
By changing the sign of the light-quark mass, the power
corrections for the vector correlator $\Pi_{\mu\nu}(p,m_Q,m_q)$
are easily found from $\Pi^5_{\mu\nu}(p,m_Q,m_q)$: $\Pi^{\rm
power}_{\mu\nu}(p,m_Q,m_q)=\Pi^{\rm 5,power}_{\mu\nu}(p,m_Q,-m_q)$. 
Obviously, the Borelized power
corrections to both the pseudoscalar and the vector correlators vanish in the limit $\tau=0$.

Since the power corrections do not contribute to the LD sum rule
under consideration, we need to consider only the perturbative contributions. After applying the duality cuts at $s_{\rm eff}$,
separately in the pseudoscalar and the vector channels, performing
the Borel transform $p^2\to\tau$, and setting $\tau\to 0$, the
corresponding sum rules take the form
\begin{eqnarray}
\label{Pia}
f_{P,V}^2&=&\int\limits_{(m_Q+m_q)^2}^{s_{\rm eff}}ds \;\rho^{\rm pert}_{P,V}(s,m_Q,m_q).
\end{eqnarray}
The functions $\rho^{\rm pert}_{P}$ and $\rho^{\rm pert}_{V}$ in (\ref{Pia}) 
are the spectral densities of the
invariant functions $\Pi^5_{L}(p^2)$ and $\Pi_{T}(p^2)$, respectively.

Let us emphasize that in (\ref{Pia}) both the full spectral
densities and the decay constants are scale independent
quantities. Therefore, the effective thresholds are
scale independent objects, too. In perturbation theory, the
spectral densities are calculated as power expansions in $a \equiv
\alpha_s(\mu)/\pi$, $\alpha_s(\mu)$ the strong coupling in the
$\overline{\rm MS}$-scheme at scale $\mu$:
\begin{equation}
\label{rhopert}
\rho^{\rm pert}_{i}(s,m_Q,m_q) = \rho_{i}^{(0)}(s,m_Q,m_q) + a\rho_{i}^{(1)}(s,m_Q,m_q) + 
                                                     a^2 \rho_{i}^{(2)}(s,m_Q,m_q) + O(a^3)
\end{equation}
with $i=P,V$. In practice, one adopts truncated expansions of the spectral
densities; this leads to a scale dependence of the spectral
densities. As the result, the effective thresholds will also
depend on the scale, to compensate the scale dependence of the
spectral densities emerging in the course of truncation.
Explicitly, the leading-order (LO) spectral densities read
\begin{eqnarray}
\label{rho}
\rho_P(s,m_Q,m)&=&\frac{N_c}{8\pi^2}\left(s-(m_Q-m_q)^2\right) (m_Q+m_q)^2\frac{\lambda^{1/2}(s,m_Q^2,m_q^2)}{s^3} \theta(s-(m_Q+m_q)^2),
\\[2mm]
\rho_V(s,m_Q,m)&=&\frac{N_c}{24\pi^2}\left(s-(m_Q-m_q)^2\right) \left(2s+(m_Q+m_q)^2\right)\frac{\lambda^{1/2}(s,m_Q^2,m_q^2)}{s^3} 
                                   \theta(s-(m_Q+m_q)^2).
\end{eqnarray}
Obviously, the lower integration limit in (\ref{Pia}) is
determined by the threshold in the correlation functions.

In Eq.~(\ref{rhopert}), we may employ different definitions of the
quark masses: The most advanced calculation of the pseudoscalar
and vector spectral densities including order-$O(a^2)$ terms was
performed \cite{chetyrkin}, for a massless light quark, in terms
of the heavy-quark pole mass. The expansion in terms of the
heavy-quark pole mass is appropriate for considering the
heavy-quark limit, which we address in Sec.~\ref{Section3.1}.

However, the pole-mass expansion leads to a rather slow
convergence of the perturbative expansion for the decay constants
\cite{jamin,lms_fp1,lms_fp2}. The convergence improves
considerably when one rearranges the perturbative expansion in
terms of the running $\overline{\rm MS}$ masses. Therefore, for
the practical analysis of the $m_q$-dependences of the meson decay
constants in Sec.~\ref{Section4}, we make use of the perturbative
expansion in terms of the running $\overline{\rm MS}$ masses of
the light and the heavy~quarks. The corresponding NLO and NNLO
functions $\rho_{P}^{(i)}$ ($i=1,2$) in (\ref{rhopert}) necessary
for such an analysis are found from the spectral densities of the
pseudoscalar correlation function given in \cite{jamin} by
multiplying them by $1/s^2$. Similarly, the transverse spectral
densities $\rho_{V}^{(i)}$ in (\ref{rhopert}) are found from the
spectral densities of \cite{khod} by multiplying them by~$1/s$. In
our analysis, we make use of the exact LO perturbative spectral
density given by (\ref{rho}), at the NLO we keep the terms $O(a
m_q^0)$ and $O(a m_q^1)$, and in the NNLO we keep the only known
terms of order $O(a^2m_q^0)$.

We would like to emphasize that the perturbative spectral density
(\ref{rhopert}) does not generate terms of order $m_q\log(m_q)$ in
the dual correlator (\ref{Pia}). This observation will be crucial
for discussing properties of the effective thresholds in~the next
section.

\section{\label{Section3}Dependence of the effective thresholds on the quark masses}

Let us now consider the dependences of the effective threshold on
the quark masses $m_Q$ and $m_q$.

\subsection{\label{Section3.1}Heavy-quark limit in the pole-mass scheme}
We start with the heavy-quark limit of the decay constants, originally 
discussed in Refs.~\cite{Neubert:1991sp,Bagan:1991sg} within the Heavy 
Quark Effective Theory (HQET).
In what follows, however, we do not consider the static decay constants and 
we work in full QCD.

For the sake of argument we consider first a massless light
quark: $m_q=0$. We can make use of any scheme for the heavy-quark
mass, but we start with the pole-mass scheme, which leads to a
more transparent behaviour of the effective threshold. We first
isolate the pole mass, which we denote $M_Q$, in the effective
threshold:
\begin{equation}
\sqrt{s_{\rm eff}}=M_Q+z^{\rm pole}_{\rm eff}(M_Q).
\end{equation}
For the decay constant of pseudoscalar and vector $\bar Qq$
mesons, using results from \cite{chetyrkin}, we obtain in the
limit $M_Q\to\infty$
\begin{eqnarray}
\label{HQlimit1}
f_{P}^2M_Q&=&\frac{N_c}{3\pi^2}(z^{\rm pole}_{\rm eff})^3\left[1+\bar a\,\frac{C_F}{12}
\left\{45+4\pi^2 + 18\log\left(M_Q/2 z^{\rm pole}_{\rm eff}\right)\right\}+O(\bar a^2)\right],
\nonumber
\\
f_{V}^2M_Q&=&\frac{N_c}{3\pi^2}(z^{\rm pole}_{\rm eff})^3\left[1+\bar a\,\frac{C_F}{12}
\left\{33+4\pi^2 + 18\log\left(M_Q/2 z^{\rm pole}_{\rm eff}\right)\right\}+O(\bar a^2)\right].
\end{eqnarray}
Hereafter, we denote $\bar a\equiv \bar \alpha_s(M_Q)/\pi$,
$\bar \alpha_s(M_Q)$ the running strong coupling in the
$\overline{\rm MS}$ scheme at scale $M_Q$, and use the standard
notations $C_F=(N_c^2-1)/(2N_c)$, $C_A=N_c$, $T=1/2$, and $n_l$
the number of massless quarks \cite{chetyrkin}.

Since only the near-threshold behaviour of the spectral densities
is relevant for the leading behaviour in the large-$M_Q$ limit, we
may obtain also the $O(\bar a^2)$ terms in the dual correlation
functions (i.e., the r.h.s.\ of (\ref{HQlimit1})) from the
analytical expressions for these spectral densities given by
Eqs.~(30) and (31) of \cite{chetyrkin}.

In the limit $M_Q\to\infty$, the dual correlation functions,
expressed in terms of $z^{\rm pole}_{\rm eff}(M_Q)$, do not
contain corrections of order $\bar a^n M_Q$ (this property will
not hold in the running-mass scheme) but still contain
$\log(M_Q)$ terms of the type $\left(\bar a \log\left(M_Q/z^{\rm
pole}_{\rm eff}\right)\right)^n$, $\bar a\left(\bar a
\log\left(M_Q/z^{\rm pole}_{\rm eff}\right)\right)^{n-1}$, etc.
The terms $\left(\bar a \log\left(M_Q/z^{\rm pole}_{\rm
eff}\right)\right)^n$, although formally of order $\bar a^n$,
remain unsuppressed in the limit $M_Q\to\infty$. To treat all
terms containing $\log(M_Q)$, it is important to emphasize that
they are exactly the same in the vector and the pseudoscalar sum
rules. Therefore, they may be resummed by introducing a properly
defined effective threshold $z^{\rm HQ}_{\rm eff}$, one and the
same in the pseudoscalar and the vector channels. The explicit relation between $z^{\rm pole}_{\rm eff}(M_Q)$ and $z^{\rm
HQ}_{\rm eff}$, including also the $a^2$ terms, reads:
\begin{eqnarray}
\label{HQlimit2} z^{\rm pole}_{\rm eff}(M_Q)&=&z^{\rm HQ}_{\rm eff} 
\left[ 1-\bar a\, d_{11}\log\left({M_Q}/{z^{\rm HQ}_{\rm eff}}\right) - \bar a^2 d_{22}\left(\log\left({M_Q}/{z^{\rm HQ}_{\rm eff}}\right)\right)^2 -
\bar a^2 d_{21}\log\left({M_Q}/{z^{\rm HQ}_{\rm eff}}\right)+O(\bar a^3)\right],\qquad
\\
d_{11}&=&\frac{C_F}{2}, \nonumber \\
d_{22}&=&\frac{C_F}{24}\left[11 C_A-3 C_F-4 n_l T\right],\nonumber
\\
d_{21}&=&\frac{C_F}{432} \left[C_F(63+48\pi^2) + C_A(1401+76\pi^2-396\log 2) - 4 n_l T(129+8\pi^2-36\log 2)\right]. \nonumber
\end{eqnarray}
The new quantity $z^{\rm HQ}_{\rm eff}$, which has the meaning of
the effective threshold in HQET, absorbs all $\log(M_Q)$ terms on
the r.h.s.\ of the sum rules (\ref{HQlimit1}); the latter assume a
form in which the HQ limit may be easily taken:
\begin{eqnarray}
\label{HQlimit1b}
f_{P}^2M_Q&=&\frac{N_c}{3\pi^2}(z^{\rm HQ}_{\rm eff})^3
\left[1+ \bar a \frac{C_F}{12}\left\{45+4\pi^2-18\log 2\right\} + O(\bar a^2)\right],
\nonumber
\\
f_{V}^2M_Q&=&\frac{N_c}{3\pi^2}(z^{\rm HQ}_{\rm eff})^3
\left[1+ \bar a \frac{C_F}{12} \left\{33+4\pi^2-18\log 2\right\} + O(\bar a^2)\right].
\end{eqnarray}
We did also calculate the $O(\bar a^2)$ contributions but do not
present their explicit expressions here. The expressions
(\ref{HQlimit1b}) immediately lead to the ratio of the decay
constants in the heavy-quark limit \cite{neubert,grozin}. Including also
$O(\bar a^2)$ corrections, we obtain\footnote{The second-order
pseudoscalar and vector spectral densities near the threshold,
Eqs.~(30) and (31) in \cite{chetyrkin}, contain three unknown
constants, $\tilde c_{FF}$, $\tilde c_{FA}$, and $\tilde c_{FL}$,
which cancel in the ratio ${f_{V}}/{f_P}$.}
\begin{eqnarray}
\label{fVoverfP}
\frac{f_{V}}{f_P}=1-\bar a\frac{C_F}{2}+\bar a^2\frac{C_F}{144}&\Big\{& 93 C_F +4(-41+19 n_l)T + C_A\left(-263-24\pi^2(\log 2-1)\right) \nonumber \\
&+&16\pi^2\left(T+C_F(\log 8 -4)\right) + 36(C_A-2 C_F)\zeta_3\Big\} + O(\bar a^3),
\end{eqnarray}
with $\zeta_3\simeq 1.202$. The $O(\bar a^2)$ term in (\ref{fVoverfP}) reproduces the result first presented in Eq (3.12) of 
\cite{grozin}. 

Notice that for finite $M_Q$, $z^{\rm pole}_{\rm eff}$ contains not only the logarithmic corrections,
which are the same in the pseudoscalar and the vector channels,
but also the $1/M_Q$ corrections,
\begin{equation}
z^{\rm pole}_{\rm eff} = z^{\rm HQ}_{\rm eff}\Big[1-\bar a\frac{C_F}{2} \log\left({M_Q}/{z^{\rm HQ}_{\rm eff}}\right) + O(\bar a^2)\Big]+O(1/M_Q),
\end{equation}
which are different for the thresholds in the pseudoscalar and the
vector sum rules.

\subsection{Combined heavy-quark and chiral limits in the pole-mass scheme}

The results (\ref{HQlimit1}) are obtained for a massless light
quark. Switching on a small light-quark mass $m_q$, the leading
corrections generated by integration of the perturbative spectral
densities are proportional to $m_q$. As already noted in \cite{lms_ib_2017}, no
chiral logs of the kind $m_q \log(m_q)$ arise from integrating the
spectral densities. Therefore, chiral logs in the decay constants
may be generated only by chiral logs in the effective threshold.
Moreover, in order to study the chiral logs in the decay
constants, it is sufficient to make use of the perturbative
spectral densities for $m_q=0$. On the other hand, heavy-meson chiral perturbation theory (HMChPT)  
\cite{sharpe}, requires the appearance of chiral logs, which we denote
as $z^{\rm HQ}_L$ in the chiral expansion of the decay constants
in the heavy-quark limit. Since the only source of such terms is
the effective threshold, we write
\begin{equation}
\label{3.6} z^{\rm HQ}_{\rm eff}=z^{\rm HQ}_{0}\left(1+z^{\rm HQ}_{L}\right)+\ldots,
\end{equation}
where the dots denote linear and higher-order terms in the
light-quark mass $m_q$. The coefficient $z^{\rm HQ}_{L}$ can now
be fixed by matching to HMChPT \cite{sharpe}, which provides the
explicit chiral logs $R_\chi(m_q)$ in the ratio $f_{H_q}(m_q)/f_{H_{ud}}$, 
with $H_{ud}$ a heavy meson with a light
valence quark of the average mass $m_{ud}\equiv(m_u+m_d)/2$.
Finally, we obtain
\begin{equation}
\label{zL} z^{\rm HQ}_L(m_q)=\left[1+R_\chi(m_q)\right]^{2/3}-1\approx \frac23 R_\chi(m_q).
\end{equation}
The explicit expression for $R_\chi(m_q)$ was derived in
\cite{sharpe} and presented by Eq.~(A.3) of \cite{lms_ib_2017}.

\subsection{Quark-mass dependences of the effective threshold in the running-mass scheme}

For practical sum-rule analyses of decay constants, one prefers
the $\overline{\rm MS}$ running-mass scheme since it entails
a~better convergence of the perturbative expansion
\cite{jamin,lms_fp1,lms_fp2}. It is not difficult to perform the
limit $\overline{m}_Q(\mu)\to\infty$ also for the running-mass
correlation function. Also therein one can write
\begin{equation}
\label{seff_3.9}
\sqrt{s_{\rm eff}}=\overline{m}_Q(\mu)+\bar{z}_{\rm eff}(\mu).
\end{equation}
The effective threshold $\bar{z}_{\rm eff}(\mu)$ in the
$\overline{\rm MS}$ scheme is related to $z^{\rm pole}_{\rm eff}$
introduced in the pole-mass scheme through an obvious relation
which just expresses the fact that the upper integration limit
$s_{\rm eff}$ is a scheme-independent quantity:
\begin{equation}
M_Q+z^{\rm pole}_{\rm eff}=\overline{m}_Q(\mu)+\bar{z}_{\rm eff}(\mu).
\end{equation}
In particular, for $\mu=\overline{m}_Q$, taking into account that
\begin{equation}
M_Q=\overline{m}_Q\left(1+C_F \bar a\right), \qquad
\overline{m}_Q\equiv \overline{m}_Q(\overline{m}_Q),
\end{equation}
one finds
\begin{equation}
\label{3.8}
\bar{z}_{\rm eff}(\overline{m}_Q)=z^{\rm pole}_{\rm eff}+C_F \bar a \overline{m}_Q+O(\bar a^2).
\end{equation}
Since $z^{\rm pole}_{\rm eff}$ does not contain terms scaling as
$M_Q$ in the limit $M_Q\to\infty$, $\bar z_{\rm eff}(\mu)$ should
contain terms which diverge as powers of $a^n M_Q$ in this limit.
This is, of course, no obstacle for using $\bar z_{\rm eff}(\mu)$
in the analysis of the decay constants~of charmed or beauty mesons
but makes this quantity not particularly convenient for studying
the heavy-quark limit~of the sum rules. The terms in $\bar{z}_{\rm
eff}(\mu)$ divergent as $m_Q\to\infty$, however, do not lead to
divergent terms in the decay constants; also, the behaviour of the
spectral densities in the $\overline{\rm MS}$ scheme is a bit more
tricky than in the pole-mass scheme. The dual correlator is
determined by the end-point behaviour of the spectral densities;
as already mentioned in \cite{jamin}, the higher-order spectral
densities in the $\overline{\rm MS}$ scheme do not vanish at the
threshold but behave as $v^{2-k}\alpha_s\log(v)^k$,
$v=\frac{1-s/M_Q^2}{1+s/M_Q^2}$. Finally, when the $\overline{\rm
MS}$ spectral densities are used and the duality cut is expressed
via $z^{\rm HQ}_{\rm eff}$, all terms containing powers of
$\overline{m}_Q$ --- those coming from the integrals of the
spectral densities and those contained in $z_{\rm
eff}(\overline{m}_Q)$ --- cancel each other, yielding a sum rule
for $f_H^2$ that can be also obtained just by expressing $M_Q$ via
$\overline{m}_Q$ in (\ref{HQlimit1}),~e.g.,
\begin{equation}
\label{HQlimit3} f_{P}^2\overline{m}_Q=
\frac{N_c}{3\pi^2}\left(z^{\rm HQ}_{\rm eff}\right)^3\left[1+\frac{1}{12}\bar a C_F(33+4\pi^2-18 \log 2)\right].
\end{equation}
Let us now switch on a small light-quark mass $m_q$. The spectral
densities are now treated as functions of $\overline{m}_Q(\mu)$
and $\overline{m}_q(\mu)$. Taking into account that the effective
threshold depends on the scale $\mu$ only because of the
truncation of the perturbative series, and that the chiral logs
have been fixed in the pole-mass scheme, it is convenient to work
with the following parameterization for $s_{\rm eff}$:
\begin{equation}
\label{seff}
\sqrt{s_{\rm eff}} = M^{(2)}_Q+z_{\rm eff}^{\rm pole}(1+z_L^{\rm HQ})+\overline{m}_q(\mu)+\bar z'_1(\mu)\overline{m}_q(\mu) + O(\overline{m}_q^2).
\end{equation}
The pole mass $M^{(2)}_Q$ here is understood as being expressed
via the running mass $\overline{m}_Q(\mu)$ (e.g.,
\cite{polerunning}) at $O(a^2)$ accuracy, the available accuracy
of the correlation function. We can rewrite this expression in a
form similar to (\ref{seff_3.9}) in terms~of
$\bar{z}_{0}(\mu)=z_{\rm eff}^{\rm pole}+\delta
\overline{m}_Q(\mu)$, where $\delta \overline{m}_Q(\mu)\equiv
M^{(2)}_Q-\overline{m}_Q(\mu)$:
\begin{equation}
\label{seffbar}
\sqrt{s_{\rm eff}} = \overline{m}_Q(\mu)+\overline{m}_q(\mu)+\bar{z}_{0}(\mu) 
\Big(1+\frac{\bar{z}_{0}(\mu)-\delta \overline{m}_Q(\mu)}{\bar{z}_{0}(\mu)}z_L^{\rm HQ} + 
\bar z_1(\mu)\overline{m}_q(\mu)\Big)+O(\overline{m}_q^2).
\end{equation}
Let us recall that the chiral logs $z_L^{\rm HQ}$ have been
calculated in the heavy-quark limit; at finite values of $m_Q$,
chiral logs receive corrections which are unknown. So we take into
account only the known leading effect of chiral logs, to study
whether or not their impact on the IB is crucial.
Two other parameters of the effective threshold~--- $z_{\rm
eff}^{\rm pole}$ and $\bar z'_1(\mu)$ if one makes use of the
parameterization (\ref{seff}), or $\bar{z}_{0}(\mu)$ and $\bar
z_1(\mu)$ if one works with (\ref{seffbar}) --- are~unknown and
will be fixed by using some external benchmark results for the
decay constants from lattice QCD. The inclusion of higher-order
terms in the light-quark mass has no impact on the decay
constants; thus, such terms are not considered.

\section{\label{Section4}Numerical analysis of the sum rules}

Now, we turn to the numerical estimates. For the relevant OPE
parameters, we adopt the following numerical~input:
\begin{eqnarray}
\label{Table:OPE} 
&&(\overline{m}_d - \overline{m}_u) (2\;{\rm GeV}) = (2.67 \pm 0.22)\;{\rm MeV} ~\mbox{\cite{FLAG}} , \nonumber \\
&&\overline{m}_{ud}(2\;{\rm GeV}) \equiv \frac{\overline{m}_u + \overline{m}_d}{2} = (3.70 \pm 0.17)\;{\rm MeV} ~\mbox{\cite{FLAG}} ,\; 
\nonumber \\ 
&&\overline{m}_s(2\;{\rm GeV}) = (93.9 \pm 1.1)\;{\rm MeV} ~\mbox{\cite{FLAG}} , 
\nonumber \\
&&\overline{m}_c(\overline{m}_c) = (1.275 \pm 0.025)\;{\rm GeV} ~ \mbox{\cite{pdg}} , \nonumber \\
&&\overline{m}_b(\overline{m}_b) = (4.247 \pm 0.034)\ {\rm GeV} ~ \mbox{\cite{lms_fp2}}, 
\nonumber \\
&&\alpha_{\rm s}(M_Z) = 0.1182 \pm 0.0012 ~ \mbox{\cite{FLAG}}.
\end{eqnarray}
We have checked that employing slightly different values of the quark masses (compatible within uncertainties  
with those in Eq.~(\ref{Table:OPE})), which have been reported in the lattice analyses of pseudoscalar mesons 
($\overline{m}_b(\overline{m}_b) = (4.190 \pm 0.021)$ GeV,  
$\overline{m}_c(\overline{m}_c) = (1.286 \pm 0.030)$ \cite{FLAG}) 
or vector mesons 
($\overline{m}_b(\overline{m}_b) = (4.26 \pm 0.10)\;{\rm GeV}$ \cite{ETMC1}, 
$\overline{m}_c(\overline{m}_c) = (1.348 \pm 0.046)\;{\rm GeV}$ \cite{ETMC2},
$\overline{m}_s(\mbox{2~GeV}) = (99.6 \pm 4.3)\;{\rm MeV}$ \cite{ETMC2}), does not 
affect our numerical estimates for the IB within the quoted uncertainties. 

We work with the effective threshold in the form (\ref{seffbar}) and consider the following three Ans\"atze:
\begin{enumerate} 
\item 
``Constant'' threshold: the $\bar z_1(\mu)$ term in the effective threshold (\ref{seffbar}) and the
chiral logs $z_L^{\rm HQ}$ are neglected; the only unknown
parameter $\bar z_{0}(\mu)$ is fixed from the lattice results for
the decay constants of the isospin-symmetric heavy mesons, with
$m_q=m_{ud}$. 
\item 
``Linear'' threshold: the chiral logs
$z_L^{\rm HQ}$ are neglected and the parameters $\bar z_{0}(\mu)$
and $\bar z_1(\mu)$ are fixed by the lattice QCD results for the
decay constants at two $m_q$ values, for the isospin-symmetric and
the strange heavy mesons. 
\item 
``Linear + log'' threshold: the
known leading chiral logs represented by $z_L^{\rm HQ}$ are
included; the parameters~$\bar z_{0}(\mu)$ and $\bar z_1(\mu)$ are
fixed from the lattice QCD results for the decay constants at two
$m_q$ values, for isospin-symmetric and strange heavy
mesons.
\end{enumerate}
As we have already noted, because of the truncation of the
perturbative expansion, the truncated spectral densities depend on
the scale $\mu$. Obviously, the parameters $\bar z_0$ and $\bar
z_1$ are also $\mu$-dependent. 

For fixing the parameters of the
effective thresholds, we make use of the following results from
lattice QCD:
\begin{eqnarray}
\label{Table:lattice} 
&&f_D=(212.15\pm 1.45)\;{\rm MeV},~ \frac{f_{D_s}}{f_D}=1.1716\pm0.0032 ~ \mbox{\cite{FLAG}} ,\nonumber
\\ 
&&f_{D^*}=(223.5\pm 8.3)\;{\rm MeV}, ~~~ \frac{f_{D^*_s}}{f_{D^*}}=1.203\pm0.054 ~\mbox{\cite{Lubicz:2017asp}} , \nonumber
\\ 
&&f_B=(186.0\pm 4.0)\;{\rm MeV}, ~~~~ \frac{f_{B_s}}{f_B}=1.205\pm 0.007 ~ \mbox{\cite{FLAG}} ,\nonumber
\\ 
&&f_{B^*}=(186.4\pm 7.1)\;{\rm MeV}, ~~~ \frac{f_{B^*_s}}{f_{B^*}}=1.197 \pm 0.055 ~ \mbox{\cite{Lubicz:2017asp}} . \nonumber 
\\
\end{eqnarray}
In these formulas, $f_H$ denotes the decay constant of the
isospin-averaged heavy--light mesons with the light-quark mass
$m_{ud}$, whereas $f_{H_s}$ denotes the decay constant of the
heavy strange mesons.

Table \ref{Table:Results} summarizes the effective thresholds
corresponding to our three Ans\"atze and presents estimates of the
strong IB effect. For our final estimates, we perform a bootstrap
analysis of the uncertainties assuming that the OPE parameters in
(\ref{Table:OPE}) have Gaussian distributions with corresponding
Gaussian errors, whereas the scale $\mu$ has a flat
distribution~in the range $1< \mu\;({\rm GeV}) <3$ for charmed
mesons and $3 < \mu\;({\rm GeV}) < 6$ for beauty mesons.

\begin{table}[htb!]
\begin{tabular}{|c|c|c|r|c|}
\hline Meson & Threshold & $z_0\;[{\rm GeV}]$ & $z_1\,[{\rm
GeV}^{-1}]\quad$ & $f_{M_d}-f_{M_u}$ \\[2mm] 
&&&&$[{\rm MeV}]$ \\ \hline
\hline 
$D$ & Constant     & $1.363\pm 0.213$ &                   & $1.222\pm 0.219$ \\
    & Linear       & $1.366\pm 0.203$ & $-0.365\pm 0.301$ & $1.050\pm 0.102$ \\
    & Linear + log & $1.225\pm 0.194$ & $-1.422\pm 0.304$ & $0.929\pm 0.088$ \\
\hline 
$D^*$ & Constant     & $1.207\pm 0.147$ &                   & $1.276\pm 0.217$\\
      & Linear       & $1.207\pm 0.138$ & $0.006\pm 0.464$  & $1.281\pm 0.389$ \\
      & Linear + log & $1.087\pm 0.139$ & $-0.978\pm 0.524$ & $1.080\pm 0.381$ \\
\hline
\hline 
$B$ & Constant      & $1.501\pm 0.143$ &                      & $0.792\pm 0.081$  \\
    & Linear        & $1.499\pm 0.134$ & $\;\,0.498\pm 0.076$ & $1.113\pm 0.108$ \\
    & Linear + log  & $1.365\pm 0.136$ & $-0.639\pm 0.147$    & $0.918\pm 0.091$ \\
\hline 
$B^*$ & Constant     & $1.534\pm 0.163$ &                      & $0.839\pm 0.076$ \\
      & Linear       & $1.533\pm 0.152$ & $0.227 \pm 0.401$    & $1.010\pm 0.317$ \\
      & Linear + log & $1.395\pm 0.152$ & $-0.938\pm 0.448$    & $0.786\pm 0.311$ \\
\hline
\end{tabular}
\caption{Parameters of the effective thresholds and resulting IB
in the decay constants of heavy pseudoscalar and vector mesons.
The parameter $z_L$ in the effective threshold for the ``linear +
log'' ansatz is fixed by ChHQET in the heavy-quark limit.}
\label{Table:Results}
\end{table}

\begin{figure}[htb!]
\begin{tabular}{cc}
\includegraphics[width=7.50cm]{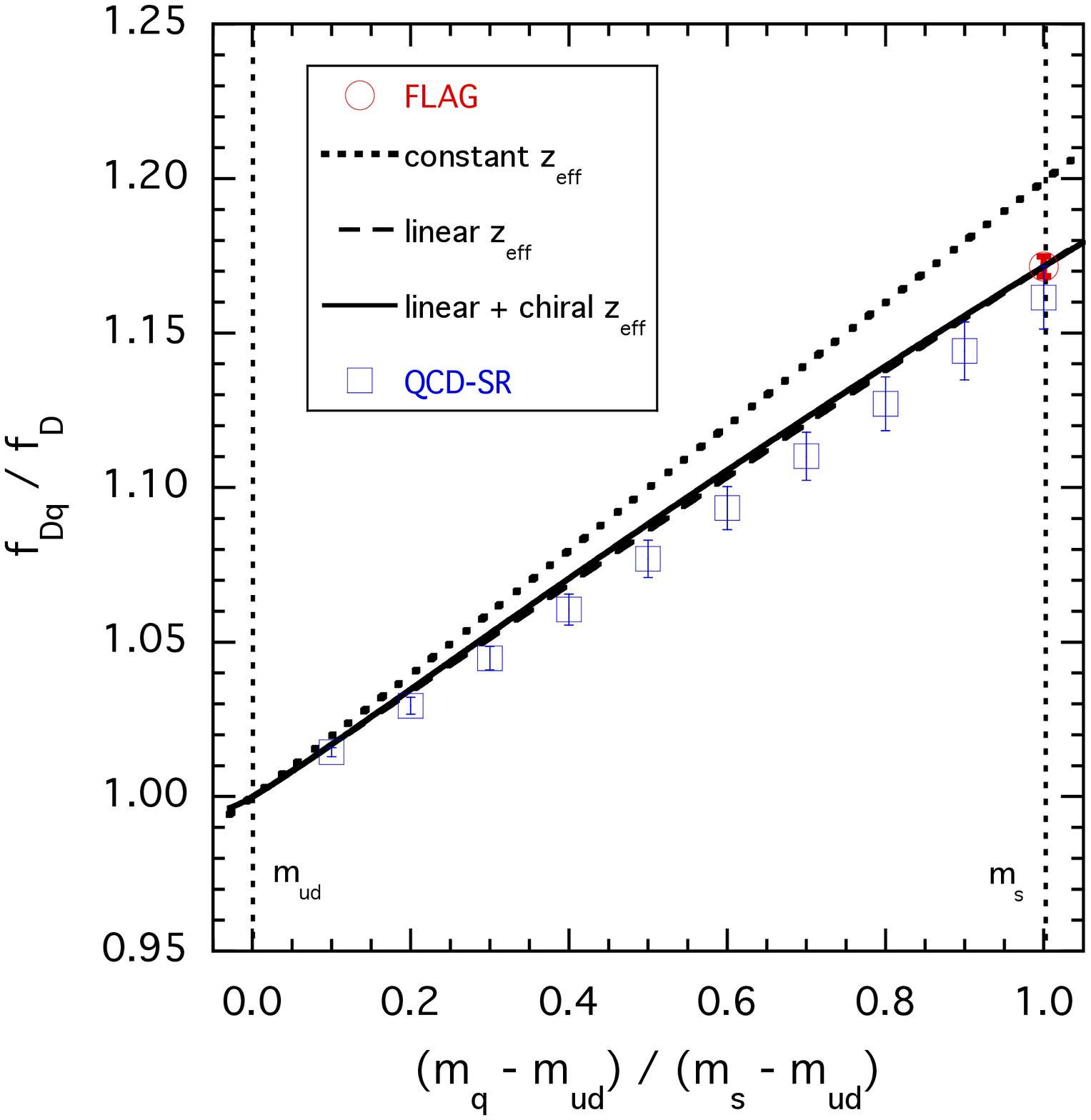}$\quad$&$\quad$\includegraphics[width=7.50cm]{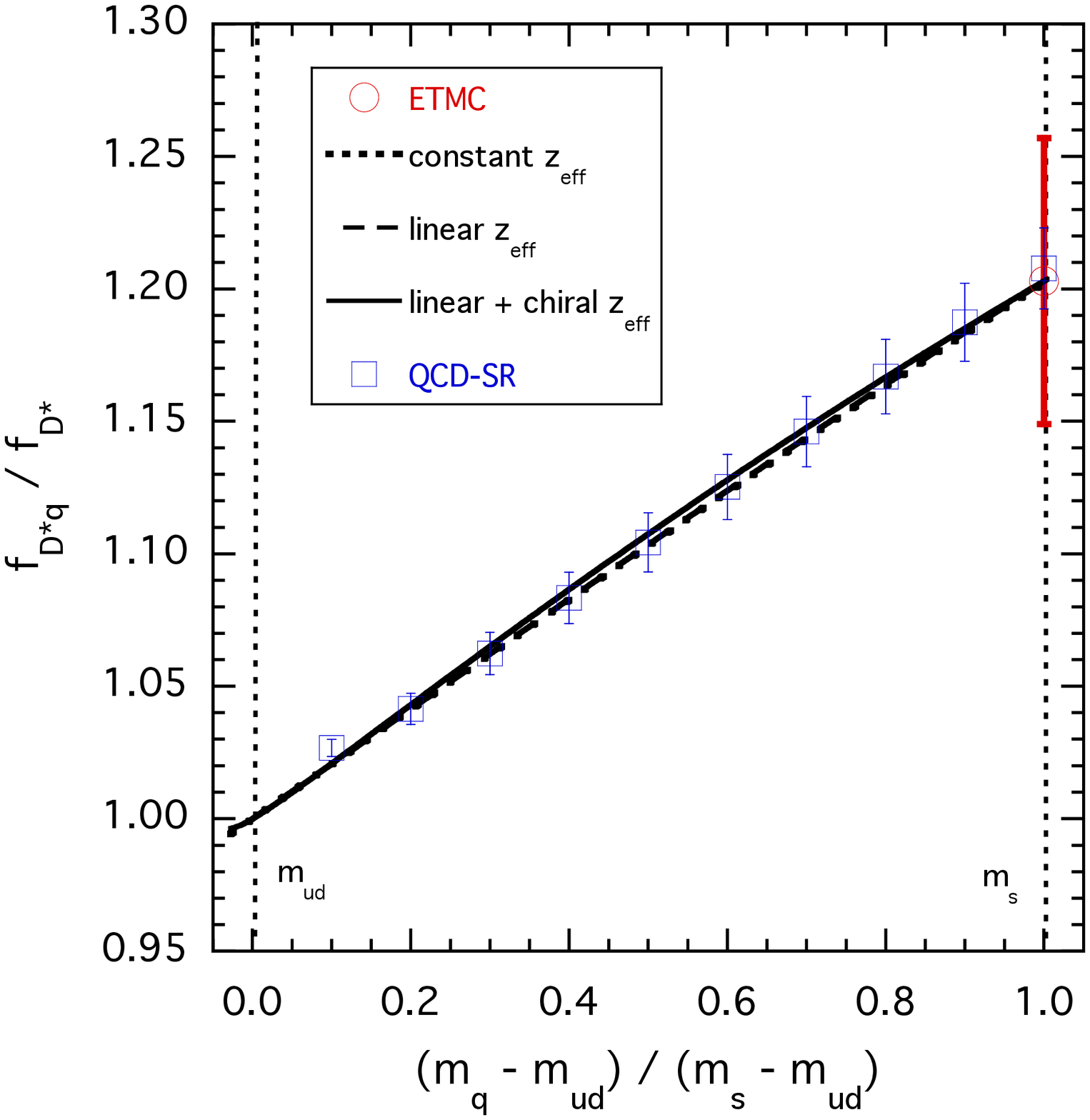}
\\(a)&(b)
\end{tabular}
\caption{\label{Plot:2} Dependence of the ratio $f(m_q)/f(m_{ud})$
for pseudoscalar $\bar c q$ (a) and vector $\bar c q$ (b) mesons. 
Dotted lines correspond to the constant ($m_q$-independent) effective threshold
[ansatz (1)] fixed from the known lattice QCD results for the
decay constants of the $D$ and $D^*$ mesons. Dashed
lines correspond to the effective threshold linear in~$m_q$
[ansatz (2)], the parameters of which are fixed by the lattice
results for strange heavy mesons $D_s$ and $D^*_s$. 
Solid lines correspond to the effective threshold
containing the known chiral logs in addition to the function
linear in $m_q$ [ansatz (3)]. In all cases, the results for the
central values of the threshold parameters in
Table~\ref{Table:Results} are displayed. We also show results from
an alternative analysis based on Borel QCD sum rules
\cite{lms_ib_2017}.}
\end{figure}

\begin{figure}[htb!]
\begin{tabular}{cc}
\includegraphics[width=7.50cm]{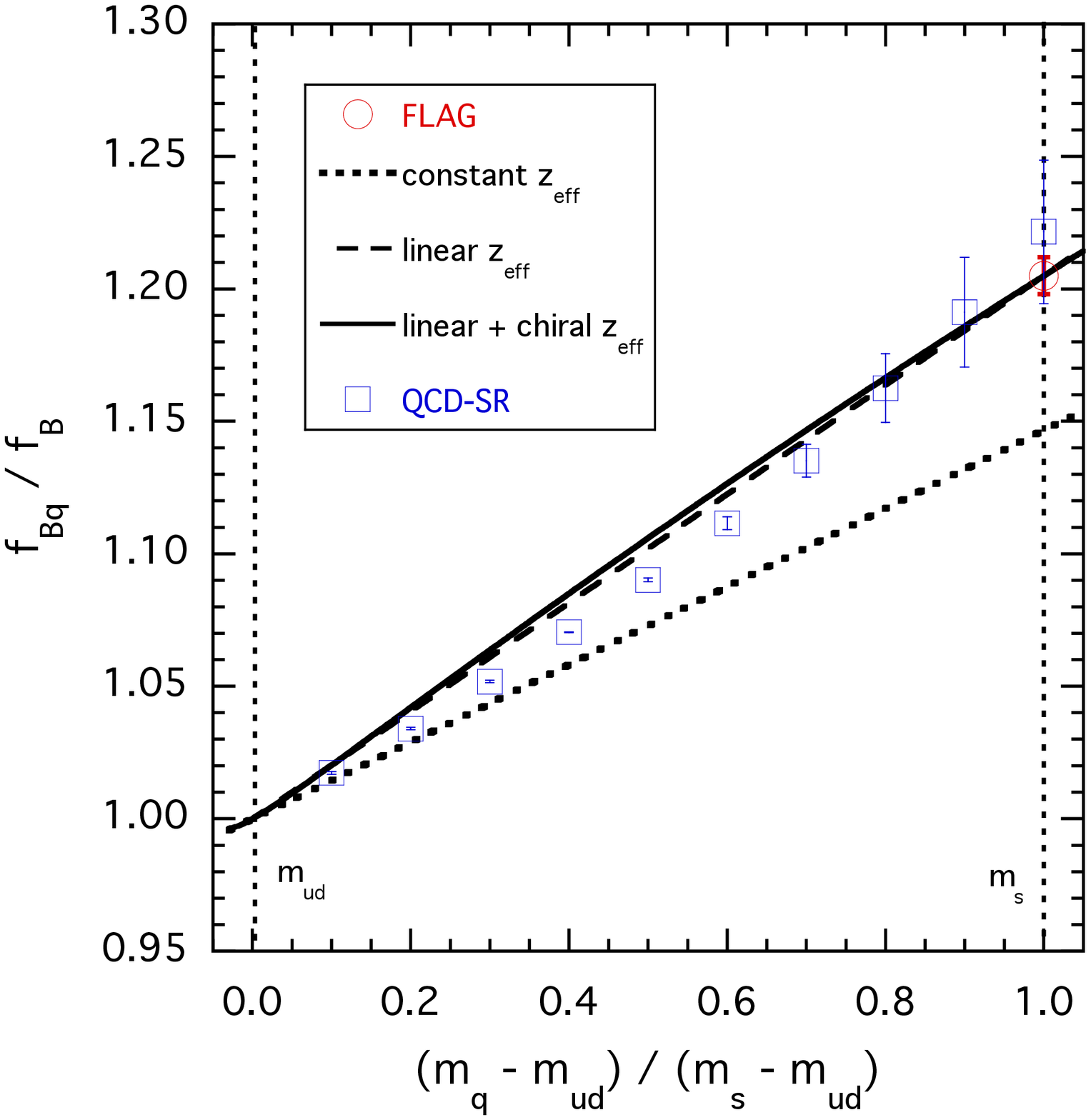}$\quad$&$\quad$\includegraphics[width=7.50cm]{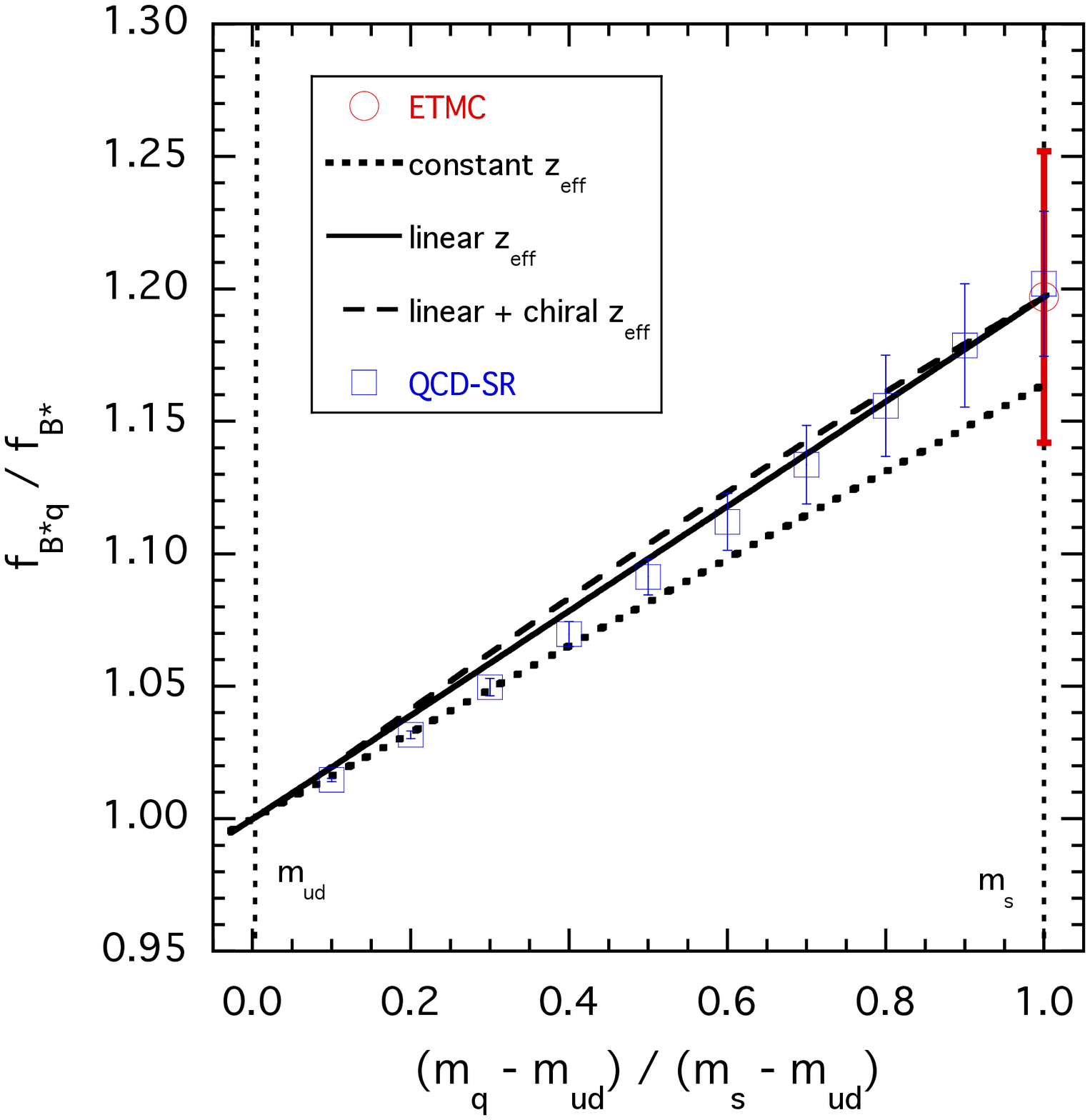}
\\(a)&(b)
\end{tabular}
\caption{\label{Plot:3} The same as in Fig.~\ref{Plot:2} but for 
pseudoscalar $\bar b q$ (a) and vector $\bar b q$ (b) mesons.}
\end{figure}

As soon as the effective thresholds are known, we readily get the
decay constants $f_{H_q}$ as a function of the scale independent ratio 
$(\overline{m}_q-\overline{m}_{ud})/(\overline{m}_s-\overline{m}_{ud})$. 
The results for the ratios of the decay constants $f_{H_q}/f_{H_{ud}}$ are shown in Figs.~\ref{Plot:2} and \ref{Plot:3}.

Notice that the results corresponding to a constant effective threshold [ansatz (1)] are quite close to 
those obtained including the $m_q$-dependence [ansatz (2)] and to the results of Ref.~\cite{lms_ib_2017}, 
which contain effects in the decay constants at any order in the light-quark mass.
So, an important conclusion to be drawn from our results is that
effects at order ${\cal{O}}(m_q^2)$ in the effective threshold are 
not crucial for describing the $m_q$-dependence of the decay
constants and for estimating the slope of the IB effect at the 
physical value of the light-quark mass: the latter are both determined 
to a large extent by the {\it known} $m_q$-dependence of the spectral
densities and can thus be reliably controlled in our approach.

\section{Summary and conclusions}

We addressed the local-duality (LD) limit, $\tau=0$, of the Borel
QCD sum rules for the decay constants of heavy--light pseudoscalar
and vector mesons. An invaluable feature of the LD limit is that
for a proper choice of the correlation function, all
vacuum-condensate contributions vanish and the full
nonperturbative QCD dynamics is parameterized~in terms of merely
one quantity --- the effective threshold. Our analysis
demonstrates that the effective threshold has~a nontrivial
functional dependence on the masses of the heavy and the light
quarks, $m_Q$ and $m_q$, respectively. This dependence has been
parameterized in the form suggested by the behaviour of the decay
constants in the known limits: the chiral limit for $m_q$ and the
heavy-quark limit for $m_Q$. In the heavy-quark limit, we clarify the role 
played by the radiative corrections in the effective threshold for reproducing 
the pQCD expansion of the decay constants of pseudoscalar and vector mesons.

This paper elucidates the dependence of the decay constants on a
light-quark mass $m_q$ in the range $m_{ud} < m_q < m_s$. Fixing a
few numerical parameters of the effective threshold by using the
available accurate inputs from lattice QCD, we have derived the
full analytic dependence of the decay constants $f_H(m_q)$ on the
light-quark mass $m_q$. The resulting dependence of the decay
constants $f_H(m_q)$ on $m_q$ emerges from two sources: (i) from
the $m_q$-dependence of the QCD perturbative spectral densities
known explicitly as expansion in powers of $\alpha_s$ and (ii)
from the $m_q$-dependence of the effective threshold known
approximately. An important outcome of our analysis is that the variation of the decay constants with respect
to $m_q$ comes to a~great extent (70--80\% of the full effect) comes
from the rigorously calculable dependence on $m_q$ of the
perturbative spectral densities and is therefore under a good
theoretical control.

Noteworthy, the known perturbative expansion of the correlation functions 
\cite{chetyrkin,jamin,khod}, where the sea-quark mass effects are neglected, limits the accuracy of the  
decay constants of the heavy-light mesons to $O(m_s \bar a^2)$ accuracy, $\bar a\sim 0.1$ at the appropriate 
renormalization scales. Therefore the accuracy of the individual decay constants obtained from QCD 
sum rules does not exceed a few MeV. Nevertheless, we would like to emphasize that the IB difference of the 
decay constants, $f_{M_d}-f_{M_u}$, where the sea-quark contributions of order $O(m_{s,u,d} \bar a^2)$ cancel each other, 
may be predicted with a much higher accuracy, $O(\delta m\bar a^2)$. Therefore, the proposed method can 
potentially provide a higher accuracy of the IB effects than other approaches. 

As our final estimates of the IB, we take the average of the
results corresponding to the linear and the linear + log effective
thresholds in Table \ref{Table:Results}:
\begin{eqnarray}
\label{deltaf_D} 
f_{D^+} - f_{D^0} & = & (0.96 \pm 0.09) ~ {\rm MeV} ~ , \\ 
\label{deltaf_Dstar} 
f_{D^{*+}} - f_{D^{*0}} & = & (1.18 \pm 0.35) ~ {\rm MeV} ~ , \\
\label{deltaf_B} 
f_{B^0} - f_{B^+} & = & (1.01 \pm 0.10) ~ {\rm MeV} ~ , \\ 
\label{deltaf_Bstar} 
f_{B^{*0}} - f_{B^{*+}} & = & (0.89 \pm 0.30) ~ {\rm MeV} ~ .
\end{eqnarray}
Sizeably larger uncertainties of the IB in the decay constants of
vector mesons compared to pseudoscalar mesons are related to
larger uncertainties of the input lattice QCD results for the
corresponding ratios $f_{H_s}/f_{H_{ud}}$.

These estimates are in good agreement with the results of our
recent analysis within a different version of QCD sum rules ---
the Borel sum rules with $\tau$-dependent threshold
\cite{lms_ib_2017}:
\begin{eqnarray}
\label{deltaf_D_std} 
f_{D^+} - f_{D^0} & = & (0.97 \pm 0.13) ~ {\rm MeV} ~ , \\ 
\label{deltaf_Dstar_std} 
f_{D^{*+}} - f_{D^{*0}} & = & (1.73 \pm 0.27) ~ {\rm MeV} ~ , \\
\label{deltaf_B_std}  
f_{B^0} - f_{B^+} & = & (0.90 \pm 0.13) ~ {\rm MeV} ~ , \\ 
\label{deltaf_Bstar_std} 
f_{B^{*0}} - f_{B^{*+}} & = & (0.81 \pm 0.11) ~ {\rm MeV} ~ .
\end{eqnarray}
The only exception is the $D^*$ case, where one observes tension
between these two sets of the results; note, however, that also
the uncertainties of these predictions are rather large.

Very recently \cite{lattice_IB_fH} a new precise determination of the strong IB effect in the 
decay constants of D- and B-mesons has been carried out by the FNAL and MILC 
lattice collaborations. 

In the charm sector their result is $f_{D^+} - f_{D^0} = 1.13 (15)$ MeV, 
which nicely agrees with our findings (\ref{deltaf_D}) and (\ref{deltaf_D_std}).
As for the bottom sector, it is shown that the available HPQCD and RBC/UKQCD 
calculations \cite{lattice_IB_fB,PDG16_fPS} overestimate significantly the strong IB effect 
because of an inappropriate use of unitary lattice points (i.e.~those having the same 
mass for valence and sea light-quarks). 
The FNAL/MILC result is $f_{B^0} - f_{B^+} = 1.12 (15)$ MeV, which is in excellent agreement with our 
findings (\ref{deltaf_B}) and (\ref{deltaf_B_std}). 

Thus, our sum-rule predictions are nicely confirmed quantitatively by lattice QCD both 
for the central values and the overall uncertainties. 
This is reassuring that the strong IB effect and its uncertainty in the decay constants of heavy-light 
mesons can be reliable and accurately estimated within the QCD sum-rule approach.

It should be emphasized that the present approach based on the
combination of OPE and a few inputs from lattice QCD potentially
has fewer theoretical uncertainties than other formulations of QCD
sum rules: first, the condensate contributions, in particular,
those of the quark condensate, which produced the main OPE error
in the decay constants, vanish in the LD limit; second, the
systematic uncertainty of the sum-rule method is now encoded in
only one quantity --- the effective threshold, which may be fixed
to good accuracy due to the use of the few accurate lattice
inputs.

Thus, QCD sum rules for the mass dimension-2 Borelized
invariant amplitudes at $\tau=0$ (i.e., an infinitely large Borel
mass parameter) provide an efficient tool for the analysis of the
dependence of decay constants (and potentially of other hadron
observables) on quark masses.

Finally, we want to mention that, besides the strong IB effect due to the up and down 
quark mass difference, there are other isospin violating effects due to electromagnetism, 
i.e.~to the difference between the up and down quark electric charges.
However, the inclusion of such electromagnetic corrections within a sum-rule approach
is not a trivial task and it requires the development of new strategies going beyond the 
traditional QCD sum-rule approaches.
In this respect it is worth mentioning a new lattice strategy~\cite{Carrasco:2015xwa} 
developed to deal with QCD+QED effects on quantities that require the cancellation 
of infrared divergences in the intermediate steps of the calculation, like, e.g., the 
decay rate of charged pseudoscalar mesons~\cite{Giusti:2017dwk}.

\section*{\bf Acknowledgments}

The authors are grateful to A.~Grozin for interesting comments. 
S.~S.~warmly thanks S.~R.~Sharpe for providing the extension of the calculation of the chiral logs of
Ref.~\cite{sharpe} to the case of $N_f = 2 + 1$ dynamical quarks. 
D.~M.~was supported by the Austrian Science Fund (FWF) under project P29028.

\appendix

\section{isospin breaking in the OPE}
\label{sec:appendix}

The two-point Green function $\Pi$ of interest is given by the
functional integral
\begin{equation}
\label{1} 
\langle T(j_1(y)j_2(0))\rangle = \frac{\int D\psi(x)D\bar\psi(x)DA_\mu(x)\; j_1(y) j_2(0)e^{i\int d^4x L(x)}} 
{\int D\psi(x)D\bar\psi(x)DA_\mu(x)e^{i\int d^4x L(x)}}, \qquad
\end{equation}
where $j_1$ and $j_2$ are (gauge-invariant) operators constructed
from quark and gluon fields, and
\begin{eqnarray}
L(x)&=&L^{(0)}(x)-\frac{1}{2}\delta m\,\bar q(x)q(x),\nonumber \\
\bar q(x)q(x) &\equiv& \bar d(x)d(x)-\bar u(x)u(x), \nonumber \\[1mm]
\delta m &\equiv& m_d-m_u.
\end{eqnarray}
Here, $L^{(0)}(x)$ is the $SU(2)$-symmetric Lagrangian describing
two equal-mass quarks, with the quark-mass term
\begin{eqnarray}
m(\bar dd +\bar uu),\qquad m\equiv\frac{1}{2}(m_d+m_u).
\end{eqnarray}
Important for our argument is that the operators $j_1$ and $j_2$
do not contain light-quark masses explicitly, although~they, of
course, contain the light-quark field operators. For instance, one
may consider
\begin{eqnarray}
j_1(x)&=&\bar q(x)\gamma_\mu\gamma_5 Q(x), \nonumber \\
j_2(x)&=&(j_1(x))^\dagger=\bar Q(x)\gamma_\nu\gamma_5 q(x), \quad q=u,d. \quad
\end{eqnarray}
After expanding Eq.~(\ref{1}) in powers of $\delta m$, one finds
\begin{equation}
\label{expansion} 
i\langle T(j_1(y)j_2(0))\rangle = i\langle T(j_1(y)j_2(0))\rangle^{(0)}-\frac{\delta m}{2}i^2 
\left\langle T\left(j_1(y)j_2(0)\int\bar q(z)q(z)
dz\right)\right\rangle^{(0)} + O\!\left(\delta m^2\right),
\end{equation}
with the superscript ``0'' indicating that the full Green functions
correspond to the $SU(2)$-symmetric QCD with two light quarks with
degenerate mass $m$. Let us emphasize the appealing feature of the
expansion (\ref{expansion}): at each order in $\delta m$, one
encounters the full Green function of the $SU(2)$-symmetric QCD.

One may expect that the power corrections in the OPE for the
three-point Green function $\Gamma$ of the $SU(2)$-symmetric QCD
are the $SU(2)$-symmetric condensates, e.g., $\langle \bar u
u\rangle=\langle \bar d d\rangle\equiv \langle \bar qq\rangle$.
However, in order to obtain the contributions of interest, we need
to perform the limit $q\to 0$. This step cannot be done easily:
the OPE for $\Gamma$ is only given by local $SU(2)$-symmetric
condensates if one keeps $q^2$ large and negative; a
straightforward extension of the known power corrections to $q\to
0$ leads to a wrong result: it is known that if one naively
extends power corrections in the vector three-point function to
$q=0$, then they do not satisfy the Ward identity (see, e.g.,
\cite{lmld}).

On the other hand, one can proceed by expanding $\Pi(p)$ in powers
of the small quark mass; then the mass derivatives emerge.
Translating the expression (\ref{expansion}) into momentum space,
to $O(\delta m)$ accuracy we obtain
\begin{eqnarray}
\label{Pi}
\Pi(p)=\Pi^{(0)}(p)-\frac{1}{2}\delta m\,\Gamma^{(0)}(p,q=0),
\end{eqnarray}
where $\Pi^{(0)}(p)$ is the full two-point function of
$SU(2)$-symmetric QCD, and $\Gamma^{(0)}(p,q=0)$ is the
three-point function of the scalar current $\bar qq$ at zero
momentum transfer, also calculated in the full $SU(2)$-symmetric
theory. Consequently, finding the leading-order $SU(2)$-breaking
effects reduces to calculating the Green functions in
$SU(2)$-symmetric QCD.

Now, consider the correlation functions
\begin{eqnarray}
\Pi(p)&=&i\int d^4y e^{i p y}\langle T(j_1(y)j_2(0))\rangle, \nonumber \\
\Gamma(p,q)&=&i^2\int d^4y d^4z e^{i p y}e^{-i q z} \langle T(j_1(y)j_2(0)\bar q(z)q(z))\rangle.
\end{eqnarray}
For the two-point function, we can write the dispersion
representation
\begin{equation}
\label{Pi1}
\Pi^{(0)}(p) = \int\limits_{(m_Q+m)^2}^\infty\frac{ds}{s-p^2}\rho(s,m_Q,m) ~ \theta(s-(m_Q+m))^2.
\end{equation}
Using the well-known relation
\begin{equation}
\label{Pi0}
\Gamma^{(0)}(p,q=0) = -\frac{\partial}{\partial m}\Pi^{(0)}(p),
\end{equation}
the three-point function at zero momentum transfer may be related
to the mass derivative of the two-point function, which then leads
to the appearance of the mass derivatives of the quark condensate.



\begin{thebibliography}{99}

\bibitem{svz}
M.~A.~Shifman, A.~I.~Vainshtein, and V.~I.~Zakharov, Nucl.~Phys.~B {\bf 147}, 385 (1979).

\bibitem{aliev}
T.~M.~Aliev and V.~L.~Eletsky, Yad.~Fiz.~{\bf 38}, 1537 (1983).

\bibitem{rubinstein}
L.~J.~Reinders, H.~R.~Rubinstein, and S.~Yazaki, Phys.~Lett.~B {\bf 103}, 63 (1981); Phys.~Rep.~{\bf 127}, 1 (1985).

\bibitem{lms_1}
W.~Lucha, D.~Melikhov, and S.~Simula, Phys.~Rev.~D {\bf 76}, 036002 (2007); 
Phys.~Lett.~B {\bf 657}, 148 (2007);
Phys.~Atom.~Nucl.\ {\bf 71}, 1461 (2008); 
Phys.~Lett.~B {\bf 671}, 445 (2009); 
D.~Melikhov, Phys.~Lett.~B {\bf 671}, 450 (2009).

\bibitem{lms_new}
W.~Lucha, D.~Melikhov, and S.~Simula, 
Phys.~Rev.~D {\bf 79}, 096011 (2009); 
J.~Phys.~G {\bf 37}, 035003 (2010);  
Phys.~Lett.~B~{\bf 687}, 48 (2010); 
Phys.~Atom.~Nucl.~{\bf 73}, 1770 (2010); 
W.~Lucha, D.~Melikhov, H.~Sazdjian, and S.~Simula, Phys.~Rev.~D~{\bf 80},~114028 (2009).

\bibitem{lms_fp1}
W.~Lucha, D.~Melikhov, and S.~Simula, 
J.~Phys.~G {\bf 38}, 105002 (2011); 
Phys.~Lett.~B {\bf 701}, 82 (2011);  
Phys.~Lett.~B {\bf 735}, 12 (2014).

\bibitem{lms_fp2}
W.~Lucha, D.~Melikhov, and S.~Simula, Phys.~Rev.~D {\bf 88}, 056011 (2013).

\bibitem{lms_fB_ratio}
W.~Lucha, D.~Melikhov, and S.~Simula, EPJ Web Conf.~{\bf 80}, 00046 (2014); 
arXiv:1411.3890; Phys.~Rev.~D {\bf 91}, 116009 (2015).

\bibitem{lms_ib_2017}
W.~Lucha, D.~Melikhov, and S.~Simula, Phys.~Lett.~B {\bf 765}, 365 (2017).

\bibitem{LD}
V.~A.~Nesterenko and A.~V.~Radyushkin, 
Phys.~Lett.~B {\bf 115}, 410 (1982); 
Phys.~Lett.~B {\bf 128}, 439 (1983). 

\bibitem{ld1}
A.~P.~Bakulev and A.~V.~Radyushkin, Phys.~Lett.~B {\bf 271}, 223 (1991); 
A.~V.~Radyushkin, Acta~Phys.~Pol.~B {\bf 26}, 2067 (1995).

\bibitem{lmld}
V.~Braguta, W.~Lucha, and D.~Melikhov, Phys.~Lett.~B {\bf 661}, 354 (2008); 
W.~Lucha and D.~Melikhov, J.~Phys.~G {\bf 39}, 045003 (2012);
I.~Balakireva, W.~Lucha, and D.~Melikhov, J.~Phys.~G {\bf 39}, 055007 (2012);
Phys.~Rev.~D {\bf 85}, 036006 (2012).

\bibitem{Lubicz:2017asp}
  V.~Lubicz {\it et al.} [ETM Collaboration],
  Phys.\ Rev.\ D {\bf 96} (2017) no.3,  034524
  [arXiv:1707.04529 [hep-lat]].

\bibitem{pdg}
C.~Patrignani {\it et al.} (Particle Data Group), Chin.~Phys.~C {\bf 40}, 100001 (2016).

\bibitem{khod}
P.~Gelhausen, A.~Khodjamirian, A.~A.~Pivovarov, and D.~Rosenthal,
Phys.~Rev.~D {\bf 88}, 014015 (2013); {\bf 89}, 099901(E)~(2014); {\bf 91}, 099901(E)~(2015).

\bibitem{chetyrkin}
K.~G.~Chetyrkin and M.~Steinhauser, 
Phys.~Lett.~B {\bf 502}, 104 (2001); Eur.~Phys.~J.~C {\bf 21}, 319 (2001).

\bibitem{jamin}
M.~Jamin and B.~O.~Lange, Phys.~Rev.~D {\bf 65}, 056005 (2002).

\bibitem{Neubert:1991sp}
  M.~Neubert,
  Phys.\ Rev.\ D {\bf 45} (1992) 2451.

\bibitem{Bagan:1991sg}
  E.~Bagan, P.~Ball, V.~M.~Braun, and H.~G.~Dosch,
  Phys.\ Lett.\ B {\bf 278} (1992) 457.

\bibitem{neubert}
M.~Neubert, Phys.~Rep.~{\bf 245}, 259 (1994).

\bibitem{grozin}
D.~J.~Broadhurst and A.~G.~Grozin, 
Phys.~Rev.~D{\bf 52}, 4082 (1995). 

\bibitem{sharpe}
S.~R.~Sharpe and Y.~Zhang,
Phys.~Rev.~D {\bf 53} (1996) 5125
 [hep-lat/9510037].

S.~R.~Sharpe, private communication, see Appendix in \cite{lms_ib_2017}.

\bibitem{polerunning}
K.~Melnikov and T.~van Ritbergen,
Phys.~Lett.~B {\bf 482}, 99 (2000).

\bibitem{FLAG}
S.~Aoki {\it et al.},
Eur.\ Phys.\ J.\ C {\bf 77} (2017) no.2,  112
 [arXiv:1607.00299 [hep-lat]].

See also S.~Aoki {\it et al.},
Eur.~Phys.~J.~C {\bf 74}, 2890 (2014)
 [arXiv:1310.8555 [hep-lat]].

\bibitem{ETMC1}
A.~Bussone {\it et al.} [ETM Collaboration],
Phys.\ Rev.\ D {\bf 93}, 114505 (2016)
 [arXiv:1603.04306 [hep-lat]].

\bibitem{ETMC2}
N.~Carrasco {\it et al.} [ETM Collaboration],
Nucl.\ Phys.\ B {\bf 887}, 19 (2014)
 [arXiv:1403.4504 [hep-lat]].

\bibitem{lattice_IB_fH}
A.~Bazavov {\it et al.} (FNAL and MILC Collaborations),
  arXiv:1712.09262 [hep-lat].

\bibitem{lattice_IB_fB}
R.~J.~Dowdall {\it et al.} (HPQCD Collaboration),
Phys.~Rev.~Lett.~{\bf 110}, 222003 (2013) [arXiv:1302.2644 [hep-lat]].

N.~H.~Christ, J.~M.~Flynn, T.~Izubuchi, T.~Kawanai, C.~Lehner, A.~Soni, R.~S.~Van de Water, and O.~Witzel,
Phys.~Rev.~D {\bf 91}, 054502 (2015)
 [arXiv:1404.4670 [hep-lat]].

\bibitem{PDG16_fPS}
J.~L.~Rosner, S.~Stone, and R.~S.~Van de Water, arXiv:1509.02220, published in \cite{pdg}.

\bibitem{Carrasco:2015xwa}
  N.~Carrasco {\it et al.},
  Phys.\ Rev.\ D {\bf 91} (2015) no.7, 074506
  [arXiv:1502.00257 [hep-lat]].
  
\bibitem{Giusti:2017dwk}
  D.~Giusti, V.~Lubicz, G.~Martinelli, C.~T.~Sachrajda, F.~Sanfilippo, S.~Simula, N.~Tantalo and C.~Tarantino,
  arXiv:1711.06537 [hep-lat].

\end{thebibliography}
\end{document}